\RequirePackage{xcolor}
\documentclass[]{rsos}

\usepackage[utf8]{inputenc}
\usepackage{float}
\usepackage{graphicx}
\usepackage{amsmath}
\usepackage{tikz}
\usepackage{breakcites}
\usepackage{lmodern}
\usepackage[T1]{fontenc}
\usepackage{tabularx}

\def\bitcoinA{%
  \leavevmode
  \vtop{\offinterlineskip 
    \setbox0=\hbox{B}%
    \setbox2=\hbox to\wd0{\hfil\hskip-.03em
    \vrule height .3ex width .15ex\hskip .08em
    \vrule height .3ex width .15ex\hfil}
    \vbox{\copy2\box0}\box2}}
    
\begin{document}

\title{Toward Open Data Blockchain Analytics: A Bitcoin Perspective}

\author{
D. McGinn
, D. McIlwraith 
and Y. Guo
}

\address{
Data Science Institute, Imperial College London\\
}

\subject{e-Science, Analysis}


\corres{Dan McGinn\\
\email{d.mcginn@imperial.ac.uk}}

\begin{abstract}
Bitcoin is the first implementation of what has become known as a `public permissionless' blockchain.  Guaranteeing security and protocol conformity through its elegant combination of cryptographic assurances and game theoretic economic incentives, it permits censorship resistant public read-write access to its append-only blockchain database without the need for any mediating central authority.  Not until its advent has such a trusted, transparent, comprehensive and granular data set of digital economic behaviours been available for public network analysis.  In this article, by translating the cumbersome binary data structure of the Bitcoin blockchain into a high fidelity graph model, we demonstrate through various analyses the often overlooked social and econometric benefits of employing such a novel open data architecture.  Specifically we show (a) how repeated patterns of transaction behaviours can be revealed to link user activity across the blockchain; (b) how newly mined bitcoin can be associated to demonstrate individual accumulations of wealth; (c) through application of the na{\"i}ve quantity theory of money that Bitcoin's disinflationary properties can be revealed and measured; and (d) how the user community can develop coordinated defences against repeated denial of service attacks on the network.  All of the aforementioned being exemplary benefits that would be lost with the closed data models of the `private permissioned' distributed ledger architectures that are dominating enterprise level development due to existing blockchain issues of governance, scalability and confidentiality.
\end{abstract}

\begin{fmtext}
\section{Introduction and Prior Work}
\label{sec:intro}

Bitcoin's release in 2009~\cite{Nakamoto2008} heralded the introduction of a novel distributed database technology that has become known as blockchain.  Reaching a fault tolerant
\end{fmtext} 
\maketitle 
\noindent consensus on state has been a well researched distributed systems problem.  But reaching such a consensus without the need for any centralized identity management system is the solution to this surprisingly overlooked problem that the invention of Bitcoin has presented.  Bitcoin's resilience relies upon public read and write access to its blockchain database where such a distributed store of publicly shared data presents many opportunities for increased access, transparency and trust without the need for any further reconciliation effort between users of the shared data.  The Bitcoin protocol specification is defined by its open-source reference implementation and its precise workings are well explained in many sources such as Bonneau et al.~\cite{Bonneau2015} or Antonopoulos~\cite{Antonopoulos2014}.  However, these pseudonymous trustless blockchain architectures as currently implemented in Bitcoin or Ethereum come with significant challenges, such as their inherent difficulty to scale and their leakage of (albeit obfuscated) private information.

In a fully trustless blockchain system, each participant must verify the activity of every other participant - an inbuilt $O(n^{2})$ scalability problem.  It is well known that the Bitcoin network currently suffers confirmation delays and becomes congested at \textasciitilde4 transactions per second (tps) and the Ethereum network becomes congested at \textasciitilde17tps (compared to Paypal which claims to handle 450tps on Cybermondays or Visa which claims to have tested up to 56,000tps).  Proposals to scale Bitcoin to these global levels in the future involve a compromise of its fully trustless nature by maintaining the original blockchain as a consolidated settlement layer only, and introducing a second layer process of off-chain transaction verification between trusted centralised counterparties known as the Lightning Network.

The leakage of private information on the Bitcoin blockchain is well studied and has, to date, focussed on the deanonimization of pseudonymous bitcoin transactions.  Reid \& Harrigan\cite{Reid2011} first took the approach of associating bitcoin address tokens to unique users by splitting the blockchain into two graph structures: a transaction network and an address network, using the former to abstract the latter into an implied user network. Both Androulaki et al.\cite{Androulaki2013} and Meiklejohn et al.\cite{Meiklejohn2013} took a similar approach by splitting the blockchain into two graphs and using the associative information leaked by the shared inputs of multi-input transactions, along with information derived from `shadow' addresses used for change amounts, to derive a consolidated graph of unique entity transactions. Ron \& Shamir\cite{Ron2013} also collapsed the transaction and address graphs into an abstract entity graph whose purpose was to explore its network properties rather than deanonimization.  The literature in this area has become notably more sparse as the data has grown to become more unwieldy.  Such established privacy deficiencies have however led to the development of more private systems such as ZCash that, whilst maintaining a public permissionless blockchain architecture, employs a zero knowledge protocol to guarantee privacy, albeit at increased computational cost.  However whilst ZCash aims for the computationally secure secrecy of shielded transactions, Qesnelle~\cite{Quesnelle2017} showed how its transactions can be associated together through behavioural patterns of usage.

To mitigate against these scaling and privacy problems, whilst also avoiding the additional resources required for zero knowledge protocols or the expensive consensus mechanisms associated with public permissionless architectures, enterprise-level blockchain solutions currently in development are gravitating towards a private permissioned blockchain model of walled-garden data-silos with access controlled by gatekeepers, as shown by the brief review shown in Table~\ref{table:ppa}.

\begin{table}[H]
\begin{minipage}{\textwidth}
\caption{At the enterprise level, there is a clear design evolution toward a private permissioned distributed ledger architecture for reasons of governance, commercial confidentiality, regulatory compliance and computational simplicity.}
\label{table:ppa}
\begin{tabular}{ p{2cm} p{11cm} }
 \hline
 Clearmatics & Limited public information although the Utility Settlement Coin project is limited to 12 members of a private consortium. \footnote{\url{https://www.clearmatics.com/utility-settlement-coin-pioneering-form-digital-cash/}} \\
 \hline
 Corda (R3) & ``Corda is designed for semi-private networks in which admission requires obtaining an identity signed by a root authority...There is no global broadcast at any point.'' \footnote{\url{https://docs.corda.net/\_static/corda-technical-whitepaper.pdf}} \\ 
 \hline
 Digital Asset Holdings & ``The Digital Asset Platform Distributed Ledger layer is a permissioned ledger accessible (for reading or writing) only by known and pre-approved parties.''\footnote{\url{http://www.the-blockchain.com/docs/Digital\%20Asset\%20Platform\%20-\%20Non-technical\%20White\%20Paper.pdf}} \\
 \hline
 Hyperledger Fabric & ``Hyperledger Fabric is a platform for distributed ledger solutions...is private and permissioned...the members of a Hyperledger Fabric network enrol through a Membership Service Provider.'' \footnote{\url{https://hyperledger-fabric.readthedocs.io/en/latest/blockchain.html\#what-is-hyperledger-fabric}} \\
 \hline
 Hyperledger Sawtooth & ``Hyperledger Sawtooth is an enterprise blockchain platform for building distributed ledger applications and networks...Sawtooth is built to solve the challenges of permissioned (private) networks.'' \footnote{\url{https://sawtooth.hyperledger.org/docs/core/releases/latest/introduction.html}} \\
 \hline
 Monax & ``Monax was the first to market with a permissionable blockchain which kick started enterprise interest...Permissioned blockchain networks differ from unpermissioned blockchain networks solely based on the presence of an access control layer built into the blockchain nodes.'' \footnote{\url{https://monax.io/use\_cases/finance/}} \\
 \hline
\end{tabular}
\end{minipage}
\end{table}

These commercial private permissioned approaches, however, negate many of the prime benefits of blockchain technology: namely the trust, transparency and socio-econometric benefits of an open data model, some of which we demonstrate here.

In this article we demonstrate the full advantage presented by the open data nature of the Bitcoin blockchain: never before has a financial transaction data set of such granularity and longevity been available for public study.  We present our exploration of this open data to develop the new field of `blockchain analytics' in order to understand dynamic behaviours within blockchain systems.  By modelling the cumbersome native blockchain data as a high fidelity graph described in Section~\ref{sec:graph}, we demonstrate how traversals of the public Bitcoin data set can derive socially useful personal and econometric information not envisaged by the original data model.  In Section~\ref{sec:viz} we make the first attempt to visualize and detect associated patterns of transactional behaviour across the entire blockchain using a path dependent query facilitated only by the adoption of the graph model we describe.  We then deploy our graph in consideration of the `coinbase' transactions of each block: transactions specially crafted by the successful miner creating new amounts of bitcoin awarded to themselves as a reward for their validation work and the method by which the bitcoin economy is inflated.  Through a combinatorial analysis of coinbase-spending transactions and their disposable extranonce bytes buried deep in the coinbase raw data, Section~\ref{sec:extranonce} shows how confidence in wealth accumulation attributed to the founder and early adopters can be increased.  In Section~\ref{sec:circulation} we develop a simple measure to explore the velocity of circulation within the bitcoin economy and examine its impact on future price moves.  We round off our set of analyses in Section~\ref{sec:spam} by demonstrating how transaction patterns associated with denial of service attacks on the bitcoin network can be identified and used by the community to defend against such attacks.

In conducting these analyses we highlight examples of the transparent benefits of the public, yet secure, open data model that blockchain technology can afford, which would be lost in any private permissioned blockchain implementation.
\section{High Fidelity Graph Model}
\label{sec:graph}

The core of the Bitcoin system is the blockchain: a continuously appended publicly distributed database storing immutable, verified records of all bitcoin transactions since system inception.  A copy of this data structure is stored and grown locally by each full network peer in a sequential series of proprietary format binary data files exemplified by the de facto reference implementation of the Bitcoin protocol.  Whilst the raw blockchain presents a complete and granular transactional data set for analysis, the binary and sequential nature of this unindexed data makes direct analysis impossible and we must look for an appropriate secondary data store informed by the structure of the data itself.  To appreciate the task at hand, an example dissection of a block of this raw binary data with its non-trivial encapsulation, lack of primary keys, implicit metadata and heterogeneous byte ordering is presented at Appendix~\ref{sec:appA}. 

We now turn to look at the relationships presented by the components of the data set.  The integrity of the blockchain is predicated upon the computational work done by the miners in solving each \textit{block}, adding upon the work having already been expended in solving the previously mined block, and each other block before it.  Thus each new block is related to each prior block in the chain.  Each valid \textit{transaction} broadcast into the system becomes related to the particular block into which it is first successfully mined.  Furthermore each transaction is composed of any number of inputs and outputs, and each \textit{input} is related to a corresponding and pre-existing unspent \textit{output} belonging to its own transaction and block, ordered at a prior point in the blockchain.  Each spending output/input pair is also necessarily related to one or more bitcoin \textit{addresses} representing the public key component of the private key required to authorise a change of ownership.  Each output records an amount of bitcoin and a cryptographic challenge expressed in an executable, stack-based \textit{script} which is required to be married to the cryptographic solution contained in a corresponding script of the input to the valid spending transaction.  There is no limit to the number of transactions that miners may decide to include in a block, but historically an arbitrary limit on the size of data in a block has applied to prevent abuse of the system (originally 32MB, reduced to 1MB in 2010, and a cap of similar order exists currently after the introduction of Segregated Witness).

\begin{center}
\begin{figure}[H]
 \includegraphics[width=0.89\textwidth]{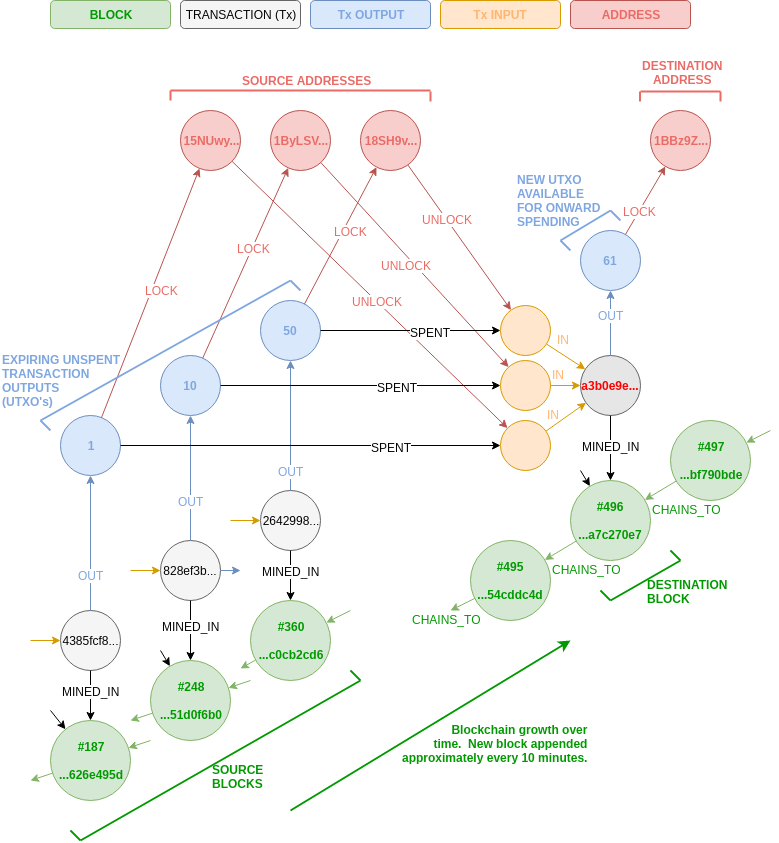}
 \caption{Example portion of the graph model of the Bitcoin blockchain showing the relationships between blocks, transactions, their inputs, outputs and associated addresses.  The figure shows the source and destination components reflecting the spending of \bitcoinA61 in the second transaction mined into Block\#496, whose identifying hash is highlighted in red.}
 \label{fig:graph}
\end{figure}
\end{center}

This unstructured tangle of data relationships between blocks, transactions, inputs, outputs and addresses naturally lends itself to a graph representation for efficient query traversal and pattern recognition.  Indeed previous work has analysed sub-graphs of the full data set, particularly with regard to the associations of identities through the abstracted relationships between addresses and transactions~\cite{Reid2011, Androulaki2013, Meiklejohn2013}.  For the purposes of knowledge discovery we propose the graph model described in Figure~\ref{fig:graph}, which refrains from abstracting information away and retains the full fidelity of the raw binary data of the blockchain whilst making for efficient query traversals that would be computationally limiting for a tabular relational database.  Also stored in our graph model (not shown in Figure~\ref{fig:graph}) are vertices to represent each of the data files and the corresponding byte offset of each block and transaction within those files for easy recourse to the raw binary data and a graphical time tree to enable temporal analyses.  The graph was implemented in the popular open source graph database Neo4j Community Edition.  

\subsection*{Implementation}

The first step in implementing the graph model was to parse the raw binary data files, each sequentially containing around 128MB of blockchain data such as that at Appendix~\ref{sec:appA}.  To this end we wrote a custom C++ parser to consume and quickly deserialize the binary data files in parallel on a 400 core HPC cluster\cite{HPC2016} into an intermediate format in preparation for import into Neo4j.

The period of the blockchain data in which we are interested is from its genesis, to shortly after the second halving of the block reward (each halving event occurring according to the Bitcoin protocol every 210,000 blocks).  Constrained by resources we arbitrarily chose to model 425,000 blocks given the second halving event occurred shortly before at Block\#420,000, therefore extracting almost 8 years of transactional data to 13 August 2016.  In our particular local copy of the blockchain, Block\#425,000 was written into the data file blk00596.dat, standing atop an accumulation of 80GB of raw data files.  The files naturally exhibit data parallelism since the blocks and transactions they contain are unique and relate to each other through unique identifiers.  It is only the address data that occurs across multiple files, and this can be rationalised in a simple post-processing step to remove data duplication.

Summary statistics of the scope of the resulting graph are shown in Table~\ref{table:stats}, which can be considered a large graph on which to compute.  The Neo4j instance was run on a 12core virtual machine with 64GB RAM, with the allocated heap space configured to sustain concurrent operations at 16GB, and 24GB allocated for the page cache.  Inevitably the memory available for the page cache results in a performance degradation due to swapping of data from disk, but to avoid this would have required a recommended 595GB of RAM for the cache to hold the entire graph resident in memory.

\begin{table}[H]
\centering
\caption{Summary statistics of vertices in the graph model.}
\label{table:stats}
\begin{tabular}{ll}
Number of Blocks          & 425,000     \\
Number of Transactions    & 148,967,063 \\
Number of Inputs          & 386,925,089 \\
Number of Outputs         & 428,714,233 \\
Number of Addresses       & 196,560,158 \\
Data size binary (MB)     & 79,924      \\
Data size Neo4j (MB)      & 519,792     \\
\end{tabular}
\end{table}

In the following sections we demonstrate how such a granular graph model can be traversed and interrogated to reveal less obvious insights into the relationships within the Bitcoin data set.
\section{The Bitcoin Blockchain: A Visual History}
\label{sec:viz}

Our aim in this section was to stress the graph database with a single query that would be forced to touch most vertices in the graph, and in so doing to create the first visualization of patterns of activity across the whole blockchain.

The query considers each input to each transaction in each block, and asks from which historical block did each input amount of bitcoin originate?  This allows us to inspect the source block (and approximately therefore the time at which it last changed hands) of each amount of bitcoin transacted in each block, which allows the examination of anomalous patterns of behaviour and as we will see in Section~\ref{sec:circulation} to explore the velocity of circulation characteristics within the Bitcoin economy.

Referring to Figure~\ref{fig:graph} it is easy to see how this query is built in the pattern matching query language `Cypher' native to Neo4j, as shown in Figure~\ref{fig:cypher}:

\begin{figure}[H]
 \includegraphics[width=\textwidth]{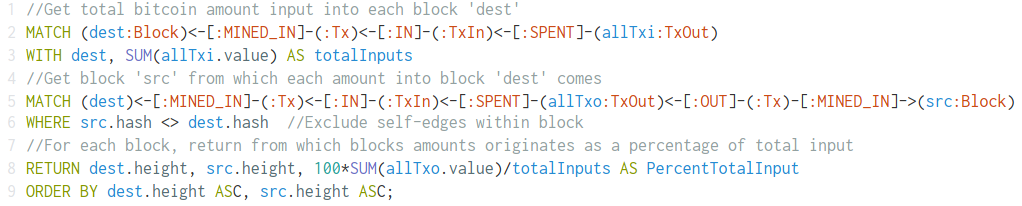}
 \caption{Cypher graph query touching all vertices to visualize source/destination flows between blocks across the entire blockchain.}
 \label{fig:cypher}
\end{figure}

In order to avoid self-edges, we do not consider newly generated coinbase transactions nor high frequency transactions whose inputs point to transactions within the same block.  We also normalize the amounts of bitcoin to be expressed as the percentage contribution to the whole amount transacted within a block in order to account for large changes in volumes transacted over time.

\begin{figure}[t]
 \includegraphics[width=\textwidth]{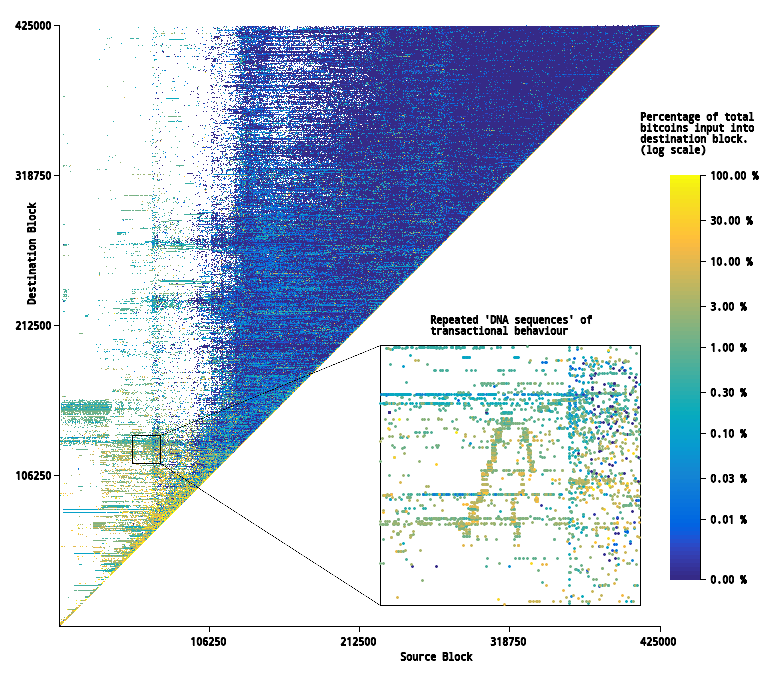}
 \caption{Full Bitcoin blockchain visualization as an adjacency matrix representation (edge-weighted by colour) of the flow of bitcoin amounts between all blocks of the entire Bitcoin blockchain to Block\#425,000, designed for interrogation on our 130 megapixel data visualization facility, a navigable interactive version of which is available at \url{https://www.doc.ic.ac.uk/~dmcginn/adjmat.html}}
 \label{fig:adjmat}
\end{figure}

Since the Bitcoin blockchain of transactions is a directed acyclic graph it naturally lends itself to an adjacency matrix representation, and since we only consider inputs from transactions prior to the current block, it takes a strictly upper triangular form.  We can now visualize this strictly upper triangular adjacency matrix, with a logarithmically coloured heat-map by the percentage contribution to each block, as shown in Figure~\ref{fig:adjmat}. 

Visualizing transfers of value between blocks on the blockchain as such an edge weighted adjacency matrix reveals several interesting features, which are better explored with the interactive zoom features mentioned in the caption to Figure~\ref{fig:adjmat}.  Primarily note the `heat' along the diagonal.  This shows that the highest percentage of value transferred into each new block originates from bitcoins that were last transacted in very recent blocks, and thus the velocity of bitcoins in circulation is high: bitcoins are predominantly transacted and churned in relatively short periods of time and this econometric feature is explored further in Section~\ref{sec:circulation}.

We can also notice distinct horizontal linear features which represent a single block \textit{into} which many previous amounts of bitcoin are consolidated and also distinct vertical features which represent a block \textit{from} which many amounts of bitcoin are distributed. Given the density of data and the space available here, these features are most notable in the sparseness of earlier blocks, particularly before the first distributive explosion around Block\#120,000.  However if we closely examine even recent blocks using our interactive tool, these horizontal and vertical patterns are still present in the data.

This allows us to speculate that transactions associated with repeated horizontal or vertical patterns in close proximity are associated behaviours, controlled by one actor.  These are highlighted as repeated `DNA sequences' of transactional behaviour in Figure~\ref{fig:adjmat} and are present throughout the blockchain.  Quantifying these relationships with a mutual similarity measure between blocks and associating transactions to incidences of high similarity are topics for future work.  We can already see though how speculative relationships from such transactions can be related to a controlling entity in our high fidelity graph to increase confidence in any transaction linkability or deanonimization tasks we may be interested to perform.

We can also employ the visualization to backtrack from a particular block on the diagonal across repeated horizontal consolidations and vertical distributions in a stepwise manner down through the blockchain to examine the primary source of such behaviour and look to correlate such anomalous behaviour with external events.
\section{Identity Leakage through Data Exhaust}
\label{sec:extranonce}

In this section we dig deep into the raw binary data, and it may be opportune to review the protocol dissection in Appendix~\ref{sec:appA} for details of a block's extranonce field.  We employ the open data derived graph database described in Section~\ref{sec:graph} to expand upon a prior primary analysis by Lerner~\cite{Lerner2013} of information leakage through parsing up to 4 potentially random functional bytes of a comment field, powerfully exposing the likely accumulation of bitcoin wealth by the very first miners, predominantly the single entity by the name of Satoshi Nakamoto~\cite{Nakamoto2008}.

As described in Bonneau~\cite{Bonneau2015}, the mechanism of Bitcoin mining is to be the first to propose to the network a block of transaction data whose summary block header has a double SHA-256 message digest that is arithmetically less than the then current difficulty criterion.  Given the nature of this problem, miners adopt a brute force approach by repeatedly testing the message digest of different block headers against the appropriate difficulty criterion.  The 80 byte block header contains 6 pieces of summary information about the set of transactions contained therein, 4 of which are fixed for any given set of immutable transaction data and network consensus.  Thus the only variables at the control of the miner in order to generate differing message digests between brute force attempts are the nonce and timestamp fields in the header, and indirectly by changing the set of transaction data, changing the set's Merkle root which is also referenced as a field in the block header (essentially a unique fingerprint of the particular ordered set of transactions contained within the block and the mechanism by which immutability is guaranteed).

Changing the transaction data set is the least preferred option since calculating its new Merkle root, validating new transactions for inclusion or removing transactions either reduce mining efficiency or reduce mining fees.  The Unix timestamp can be changed within bounds approximately -1/+2 hours of the current time, but the obvious field to test against is the 4-byte nonce field dedicated for this purpose.  However given the solution space is therefore limited to $2^{32}$ possible message digests, it is feasible that all possibilities can be exhausted by brute force within a very short period of time, where a block header solution that satisfies the difficulty criterion is not found.

To overcome this limitation it has been customary since the genesis block for miners to include up to four bytes of `extranonce' data in the redundant input field to their coinbase transaction of each block (since this first transaction of each block represents newly generated bitcoins awarded to the miner and requires no inputs).  This extranonce complements the primary nonce data of the block header.  By changing this arbitrary little-endian data included within this free-form comment field outside of the formal protocol, the whole transaction set's Merkle root is changed yet all transactions remain valid and thus a new round of $2^{32}$ attempts can be made at finding a solution to the block header.  If a miner simply increments the extranonce on the overflow of each round of $2^{32}$ unsuccessful primary nonce solution attempts, then once published in a block on success, the extranonce can be considered to represent a slow real-time clock signal from that particular miner.

\begin{figure}[H]
 \includegraphics[width=\textwidth]{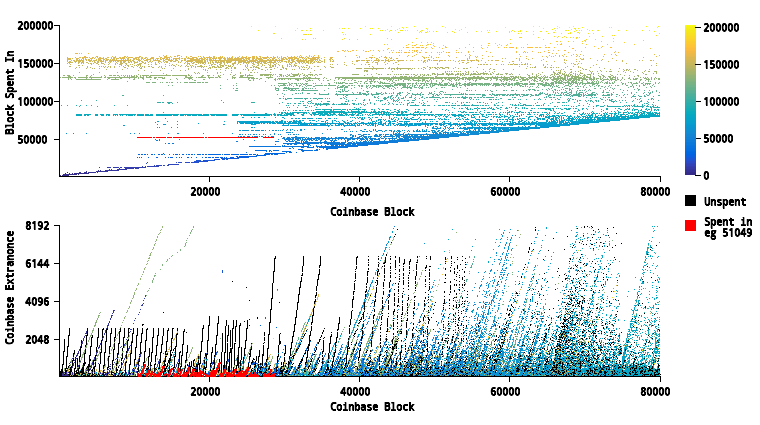}
 \caption{Plots showing heights at which each block's coinbase was first spent (top) and the extranonce value used (bottom), coloured by spent height (including unspent).  Note the constant gradient incremental extranonce features identifying discrete continuous mining operations, highlighted in red when combined with simultaneous spending data.}
 \label{fig:extranonce}
\end{figure}

The observation that a seemingly unimportant four bytes of incremental extranonce data in the general exhaust of operation actually represents a slow real-time clock of a particular miner's operation is the foundation of Lerner's 2013 analysis.  It was shown that the value of each block's incremental extranonce against its time of mining (assuming constant computational mining power) should result in a constant gradient relationship indicative of a particular miner.  Figure~\ref{fig:extranonce} replicates and expands upon Lerner's work, the bottom half showing the same obvious straight line relationships of blocks mined by particular miners, infrequently resetting the extranonce to 0.  Blocks mined by a particular miner using an infrequently resetting, non-randomised extranonce all lie on the same positively sloping line.  The slope (assuming all miners are searching the same primary $2^{32}$ nonce space) is indicative of the rate of successful block solutions, a direct measure of computational mining power and another signature of associated identity.

We extend Lerner's analysis and add to it with a traversal of the graph model to show in which block the generated bitcoins under consideration were first spent, and colour the points according to this block height (top of Figure~\ref{fig:extranonce}).  This reveals further patterns of identity associations, namely the obvious difference between spent and unspent coinbase transactions and those coinbase transactions which were all spent at the same time.  In this way even miners of low computational power with sparse extranonce data points that don't extend into a straight line above the extranonce noise can be further associated together by the linear consolidation of coinbase transactions being spent at the same time (see example highlighted in red in Figure~\ref{fig:extranonce}).

There are four reasons that we terminate this analysis so early at Block\#80,000:
\begin{itemize}
 \item{The linear features are less clear to demonstrate at larger scale given limited space here.}
 \item{More miners over time result in progressively noisier incremental data.}
 \item{The free-form nature of the coinbase field becomes more difficult to parse as participants diverge from the initial reference schema (e.g. one can witness similar incremental behaviours using sequential quotations from the Bible, which whilst leaking the same information are significantly more difficult to plot than integers!)}
 \item{Awareness of this information leakage encouraged adoption of extranonce randomization.}
\end{itemize}
\section{Monetary Supply Disinflation and Velocity of Circulation}
\label{sec:circulation}

It is well known that bitcoins come into existence as a reward to miners for ensuring system integrity through competing in the mining puzzle.  Coin creation is famously at a geometrically reducing rate, halving every four years from an initial reward of \bitcoinA50 per block such that mining rewards will cease when the amount reaches the smallest divisible unit around the year 2140, at which point approximately 21 million bitcoin will be in existence.  In some quarters this fixed algorithmic coin creation has long been a major attraction of Bitcoin as a unit of money compared to the fiat monetary system where money supply is according to the political whim of central banks.  However it has long been argued this lack of monetary expansion can be considered deflationary,~\cite{Economist2014} since expectations of a rise in value due to restricted supply will lead to hoarding.  Having every historical transaction available for scrutiny through the open data nature of the public Bitcoin blockchain allows us to examine this disinflationary claim.

Monetary economists such as Irving Fisher encapsulated this deflationary concept in the equation of exchange~\cite{Fisher1911} expounded in the na{\"i}ve quantity theory of money:

\begin{equation}
\label{eq:exchange}
MV = PT
\end{equation}

\noindent wherein changes in the monetary supply ($M$), such as the coin generation halvings programmed into Bitcoin or the contrary operation of central bank quantitative easing,  will have a causal effect primarily on the level of prices ($P$) if the velocity of circulation ($V$)  and the number of transactions ($T$) remain constant.

In the real economy, $V$ and $T$ are difficult to measure:  $V$ is often assumed fixed and $T$ is often substituted by a macro measure of national income.  However now, for the first time, the granular open data of the entire blockchain powerfully allows us to directly apply such theories against every transaction in the Bitcoin economy.

\begin{figure}[H]
 \begin{tikzpicture}[scale=0.2]
  \draw[line width=1mm] (0,1) node[anchor=north east]{$\ B_{0} $} -- 
 						(49.6,1) node[anchor=north west]{$\ B_{N}=496$};
  \draw[line width=0.5mm,->] (18.7,1.1) --
 							(18.7,1.0) node[anchor=south east]{$\ \genfrac{}{}{0pt}{}{a_{1}=1}{b_{1}=187}$};
  \draw[line width=0.5mm,->] (24.8,2.0) -- 
 							(24.8,1.0) node[anchor=south east]{$\ \genfrac{}{}{0pt}{}{a_{2}=10}{b_{2}=248}$};
  \draw[line width=0.5mm,->] (36.0,6.0) -- 
 							(36.0,1.0) node[anchor=south west]{$\ \genfrac{}{}{0pt}{}{a_{3}=50}{b_{3}=360}$};
  \fill[black] (33,-1) -- 
 			(33.8803,1) -- 
            (34.7606,-1) -- 
            cycle;
  \draw[line width=0.5mm,->] (49.6,-1.75) -- 
 							node[below]{$\ D_{496}=157.197$ blocks} (33.8803,-1.75);
 \end{tikzpicture}
 \caption{Example bitcoin dwell time measure, $D_{496}$, for the three component input amounts to all transactions mined in Block\#496. (See Figure~\ref{fig:graph} for further details)}
  \label{fig:dwelltime}
\end{figure}
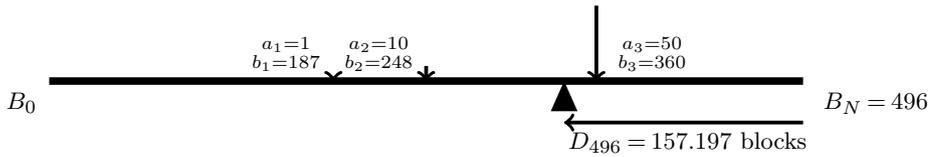

We can see from studying the diagonal of the edge weighted adjacency matrix of amounts flowing between blocks in Figure~\ref{fig:adjmat} that a large number of inputs into a block have been transacted in the very recent past, so we start by introducing a measure of velocity of circulation.  For all $m$ inputs into all the transactions mined in a particular block $B_{N}$, we define the block's bitcoin dwell time $D_{N}$ as in Equation~\ref{eq:dwelltime}:

\begin{equation}
\label{eq:dwelltime}
 D_{N}=\frac {\sum\limits_{i=1}^{m} \left ( B_{N} - b_{i} \right ) a_{i} }
             {\sum\limits_{i=1}^{m} a_{i} }
\end{equation}

\noindent where, $a_{i}$ is the amount of the input and $b_{i}$ is the block number from where the amount originates.  The dwell time can be considered the equilibrium point in time, measured in number of blocks ago, such that the weighted amount of bitcoins transacted in a block balances the imaginary beam depicted in Figure~\ref{fig:dwelltime}.  It may not be immediately obvious but these beams are the physical analogue of each row of data visualized in Figure~\ref{fig:adjmat}.

\begin{figure}[H]
 \includegraphics[width=\textwidth]{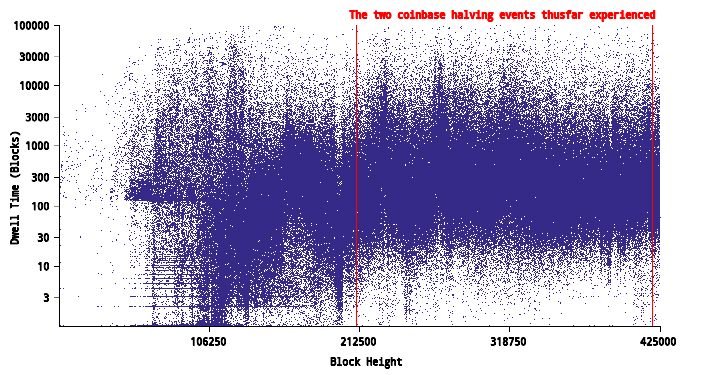}
 \caption{Bitcoin dwell time by block (log plot).}
 \label{fig:velocity}
\end{figure}

This dwell time measure is naturally inversely related to the velocity of circulation: the larger a block's dwell time, the longer transacted bitcoins in that block have been stationary and out of circulation.  Now if we look at (the log of) this dwell time measure over the entire blockchain under consideration (Figure~\ref{fig:velocity}) we can see that as volumes have increased, the velocity of circulation has reasonable variance but exhibits no accelerating or decelerating trend.

As an aside, the horizontal features observable at the beginning of the chart occur during a period of very low volume, where many blocks had single transactions of fixed amounts from a small number of blocks prior.  In fact the restriction for a miner to wait more than 100 blocks to spend the \bitcoinA50 coinbase reward can also clearly be seen in this early period.

Figure~\ref{fig:velocity} shows that of the bitcoins transacted, there is no evidence of change in any hoarding behaviour since the velocity of circulation as measured by the bitcoin dwell time also shows no significant change, despite the two halvings of the monetary supply already experienced.  In fact if we make a linear least squares fit of the dwell time data, whilst mildly positively sloping, it only increases by 33 blocks over the whole 425,000 block period from $D_{0}$=861 blocks.  Alternatively we can say that a typical bitcoin circulating over this 7-8 year period under consideration has been transacted approximately every six days (given the mining rate at \textasciitilde144 blocks/day), and this has remained the case since its genesis and over the two halving events experienced so far.  With this constant $V$ in mind, referring back to the na{\"i}ve equation of exchange~\ref{eq:exchange} as a cartoon example and noting both that the number of transactions ($T$) is constrained as a constant by the currently exhibited protocol block size limit of around 1MB and that the monetary supply ($M$) is reducing as programmed by the quadrennial halving events, we can thus theorize that the prices of goods denominated in bitcoin ($P$) must also tend to decrease from current expectations in line with with the reduced monetary supply.

Have we in fact experienced this anticipated price deflation?  Empirically there are very few goods denominated in bitcoin, but we can turn to the inverse of price and look at the purchasing power of one bitcoin.  If we look to the exchange price of a bitcoin as a measure of its purchasing power we can see it has indeed increased over the two halvings.  As well as through speculation, Bitcoin has exhibited a deflationary profile, and we can speculate that such purchasing power increases may continue as the supply becomes ever more restricted whilst the velocity of circulation and effective 1MB block size limit remain in effect.

It is only through the open data nature of the blockchain that anyone can generate on a per transaction granular basis such a metric for the velocity of circulation within the Bitcoin economy.
\section{Denial of Service Attacks}
\label{sec:spam}

\begin{figure}[H]
 \centering
 \includegraphics[width=0.5\textwidth]{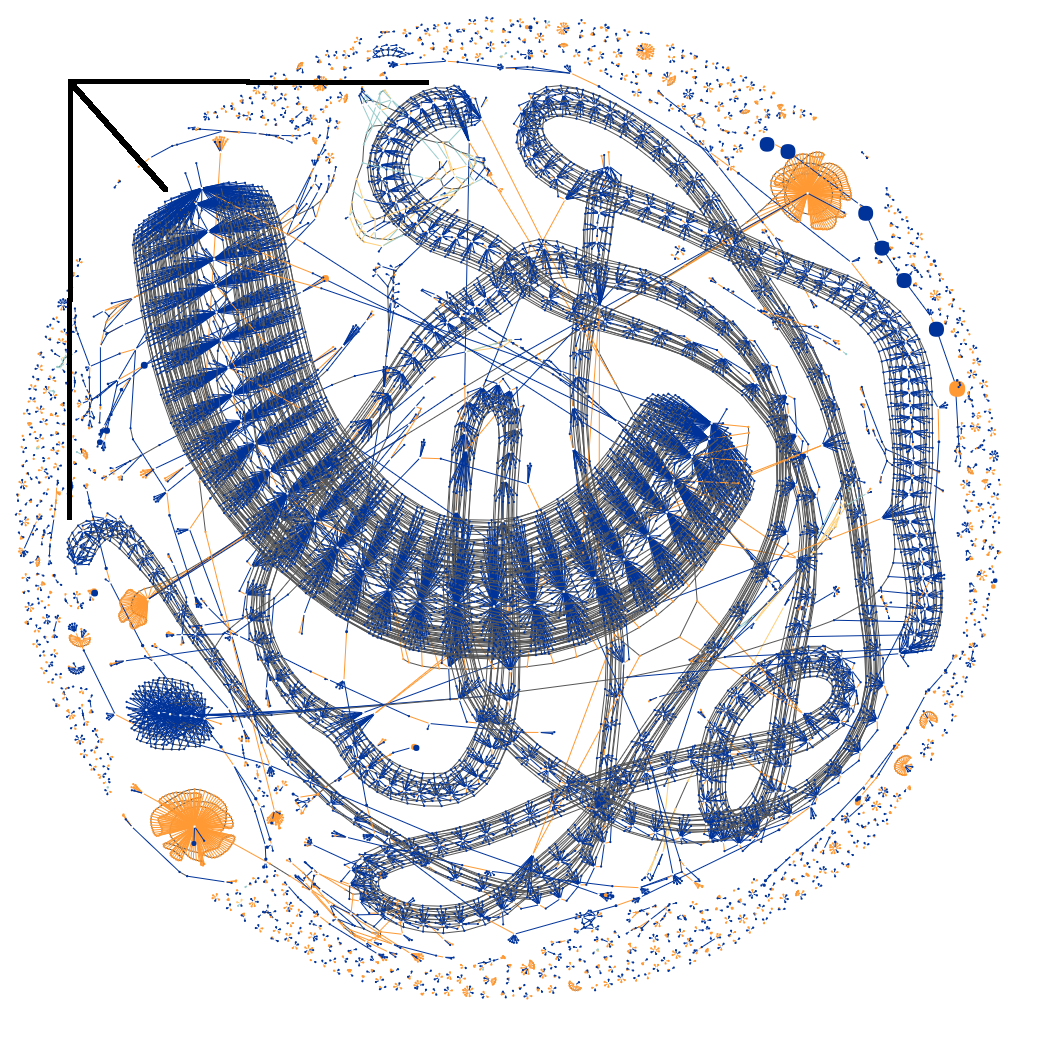}
 \caption{Algorithmically associated spam transactions forming the three visually anomalous `worm' structures (indicated) of a DoS attack commencing in Block\#364133.  (McGinn et al.~\cite{McGinn2016} for details.) }
 \label{fig:364133}
\end{figure}

In our previous work visualizing transaction patterns across the Bitcoin blockchain~\cite{McGinn2016}, a particular result was the identification of programmatically generated spam transactions.  In that work we generated a real-time force directed graph of bitcoin transactions within blocks, visualizing the relationships between transaction inputs (orange), transaction outputs (blue) and transaction components sharing a common bitcoin address (grey).  An example of the visualization showing Block\#364133 is shown in Figure~\ref{fig:364133}.  This block was mined in a period where an attacker had mounted a denial of service (DoS) attack on Bitcoin, algorithmically and cheaply generating many `spam' transactions of small value to artificially fill up the blocks with large amounts of data to push against the arbitrary the 1MB block ceiling hard coded into the Bitcoin protocol at that time.

A cursory inspection of the anomalous worm structures in the block visualization reveals the nature of the algorithm used to generate the spam transactions, namely many high frequency transactions repeatedly spending small blue outputs to 102 separate addresses in the case of the `fat worm', and 11 and 15 separate addresses in the case of the two `thin worms'.

\begin{figure}[t]
 \includegraphics[width=\textwidth]{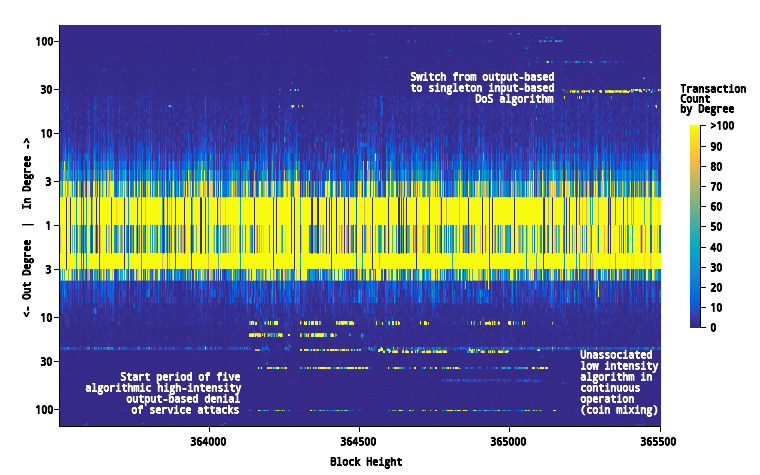}
 \caption{Spectrograph-type plot of transaction count per block by (log) +in/-out degree exposing specific periods of anomalous degree distribution.}
 \label{fig:spam}
\end{figure}

Knowing the primary feature of these high frequency spam transactions is their high out degree (and later their in degree), we can deploy our high fidelity graph model to query and explore this algorithm's evolution.  The results of this query are shown in Figure~\ref{fig:spam} which depicts, for each block, a particular `heat' according to the number of transactions of a certain (positive) in degree and the number of transactions of a certain (negative) out degree, plotted on a log scale.  Clearly there are many regular transactions in each block of small in and out degree hence the indeterminate and unimportant `heat' along the central axis, but by observing the clear linear structures away from the central axis it immediately becomes clear when the high frequency algorithm of anomalously large in or out degree is in operation. 

We can see from Figure~\ref{fig:spam} that the denial of service algorithm did indeed commence operation at the time of Block\#364133 visualized in Figure~\ref{fig:364133} and quickly evolved to generating at least 5 discrete structures, each with a unique but consistent out degree signature.  We can also see somewhat coordinated periods where the algorithms briefly cease operation, perhaps to recharge funds or perhaps for overnight shutdown which could identify the timezone of the operator.  Clearly we are able to associate these transactions by their consistent anomalous structure and coordinated start/stop behaviour.  We can also determine the point at which the algorithm ceased operation around Block\#365149, almost exactly 7 days after its start.  Shortly after this transaction output algorithm's cessation a new linear structure above the central axis emerges, indicating a high-frequency algorithm similar in nature but instead using transaction inputs as opposed to outputs, potentially collecting the small amounts that the previous algorithm had distributed, minimising the cost of the attack but multiplying its effects on the network.  It can be no coincidence that such anomalous behaviour starts so soon after the cessation of the previous pernicious algorithmic behaviour.

Another approach  performed by Baquer et al.~\cite{Baquer2016} is to use a k-means clustering of spam features.  By combining approaches we would be able to increase confidence in the positive identification of transactions related to the denial of service spam attack through the public nature of the blockchain data, and can decorate our graph model with this additional intelligence in order to identify addresses and behaviours which would otherwise remain hidden in the data and can potentially be used by the community to generate heuristic defences against such attacks.
\section{Conclusion \& Further Work}
\label{sec:conclusion}

A distributed public permissionless blockchain database such as Bitcoin securely holds immutable records of transactional data between users.  The Bitcoin blockchain is an unwieldy data structure, large in size, lacking primary keys, of non-trivial encapsulation and heterogeneous byte ordering, whilst demanding onward computation of inferred data.  In this article we have disentangled the cumbersome binary blockchain into a usable graph model with its associated benefits of efficient path traversal and pattern matching isomorphisms.  This work has taken full advantage of this first example of a granular financial open data set to show some of the socially useful analyses that can be conducted to the benefit of the system and its community of users.

Our contributions have been:
\begin{itemize}
 \item{to reveal observable patterns of linked transactional behaviour by traversing the entire graph, producing the first visualization of the entire blockchain as an edge weighted adjacency matrix.}
 \item{to increase confidence in the attribution of mined bitcoin to single entities by combining existing analyses of the extranonce with a traversal of the graph to the point at which they were first spent.}
 \item{to show econometrically a per transaction application of the quantity theory of money to the deflationary Bitcoin economy without having to rely on traditional broad and aggregative assumptions.}
 \item{to demonstrate how network metrics can distinguish anomalous patterns of algorithmically generated transaction behaviour during denial of service `spam' attacks.}
\end{itemize}

We can now set to the task of automatically classifying these linked transactional behaviours observed in the Bitcoin blockchain and decorate our graph with this additional intelligence.  We speculate that associating these transactions at the user level may reveal new patterns in the data.  We also look to apply the methods here to alternative blockchain databases such as Ethereum and ZCash, developing cross-chain analytic tools.  This has application in fields such as fraud and tax investigation, the application of econometric and economic behaviour theory and the improvement of blockchain technology in general. 

However, it must be noted that a public open data architecture such as currently implemented in Bitcoin presents challenges of privacy and scalability.  Particularly at the enterprise level, where all participants in a business network are required to be authenticated or a centralised third party can be trusted, the issues of scale and confidentiality with these distributed ledger technologies are being addressed by implementing walled-garden models of siloed data.  In such cases, however, the benefits of these private permissioned distributed ledger solutions over a properly authenticated, replicated and audited traditional database remain uncertain.  It is clear though that the open data model of a public permissionless blockchain architecture presents many often overlooked opportunities to realize additional information and value.
\clearpage
\appendix
\section{Dissecting the Raw Binary Data: Block\#170}
\label{sec:appA}

\begin{center}
\begin{sideways}
\begin{minipage}{1\linewidth}
 \hspace{-8.7cm}
 \includegraphics[scale=0.81]{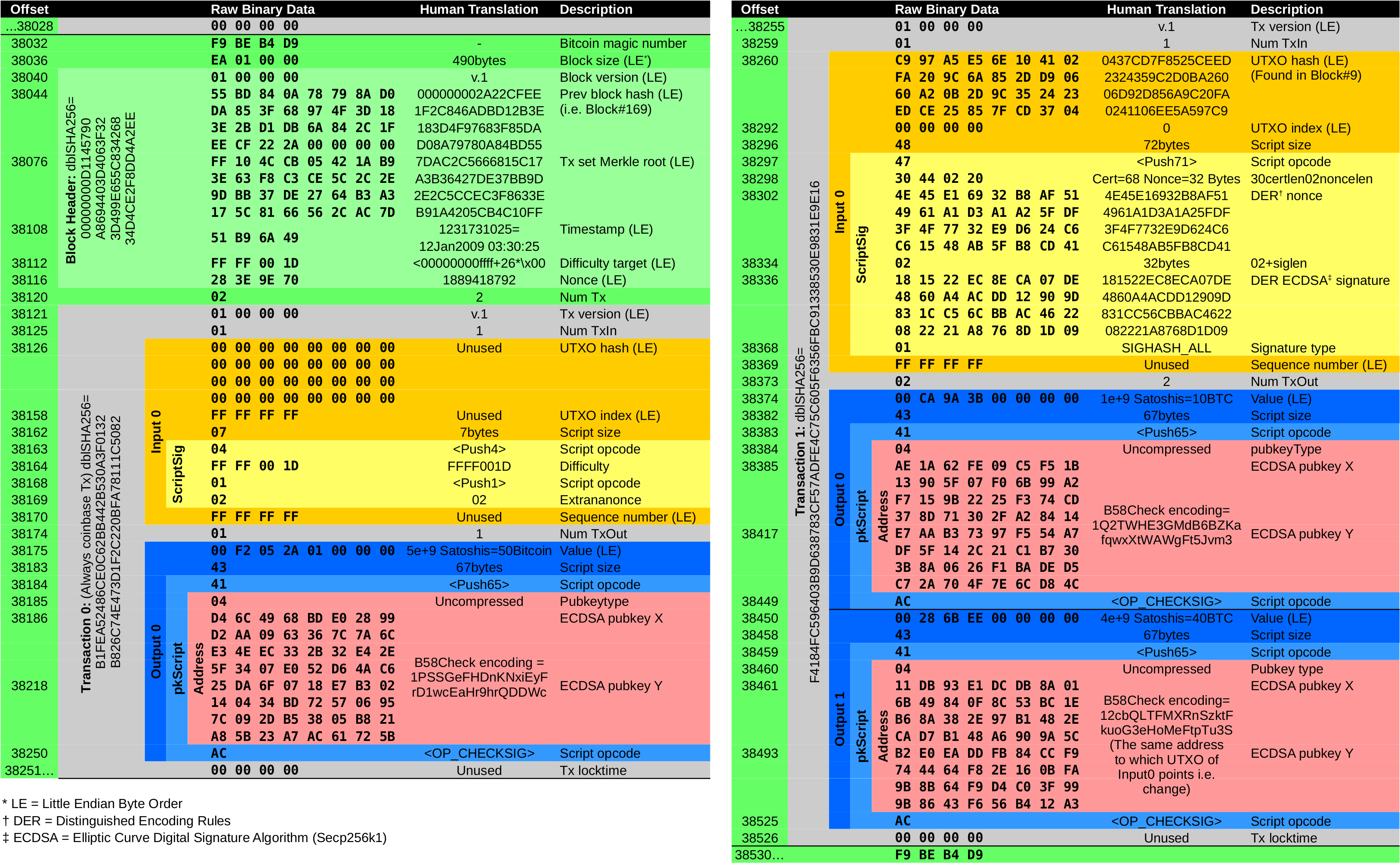}
 \label{fig:block_dissection}
 \end{minipage}
\end{sideways}
\end{center}

\clearpage
\noindent The graphic presented in this appendix shows the dissection of the 490 bytes of raw binary blockchain data representing Block\#170: the block into which the first ever spending transaction between Satoshi Nakamoto and Hal Finney was mined.  Every node fully participating in the Bitcoin network carries this data, along with that of every other valid block, leading to the robust redundancy and replication for which Bitcoin is known.

Following the colour convention of Figure~\ref{fig:graph}, the block data with its 80byte header is shown in green.  Encapsulated within the block in this case are two transactions depicted in grey.  The first transaction of a block is always the coinbase transaction allocating newly minted bitcoins to the successful miner.  Each transaction is composed of orange inputs (redundant in the case of each coinbase transaction) and blue outputs.  The transaction outputs are the sockets into which future inputs can connect, at which point they become spent.  The outputs propose an amount of bitcoin and a cryptographic challenge (pkScript) usually referencing an ECDSA public key or its derivative hash from which the bitcoin address is further derived, shown in red.  When the output's pkScript is combined with its corresponding spending input's scriptSig (which contains a DER encoded ECDSA signature over the transaction), each validating node can independently verify that the spend was correctly authorized by knowledge of the private key corresponding to the bitcoin address.

Particular to note is the over-normalization and lack of any primary keys directly identifying any piece of data such as a block hash, transaction hash or bitcoin address, all to which the data refers but must be derived from the data itself.  This fact, coupled with the non-trivial encapsulation of data of variable length and heterogeneous byte orders, leads to the result that it is necessary to post-process the binary data files, parsing them in their entirety to extract useful information.  The need for data expansion into a secondary store for efficient query traversal becomes obvious, and a graph structure is the natural choice.

It is curious to note that such great efforts were made in the Bitcoin protocol design to minimise the bytes that would have to be stored in perpetuity by every fully validating participant.  Yet the fact is that it is redundant to record the public key itself on the blockchain since ECDSA facilitates public key recovery given the signature, plain text message and nonce used, which suggests the designer(s) weren't fully conversant with elliptic curve cryptography at the outset.  An interesting aside is the ECDSA signature nonce must be truly random for any address re-use as private key recovery is possible given known public keys, signatures, corresponding plain texts and nonces~\cite{Courtois2014}.  Once linkability is established between a user's transactions through means such as those exhibited throughout this article, patterns of ECDSA nonce use may expose a feasible attack vector.

\vskip1pc

\ethics{Not applicable.}

\dataccess{This article has no additional data.  All data sourced is open data, accessible through participation in the public Bitcoin network with full integrity, consistency and liveness guarantees.  Access to the refined graph database is resource constrained but available on request to the correspondent author.}

\aucontribute{Dan McGinn conceived of the study, carried out the data collection, analysis and interpretation and drafted the manuscript; Yike Guo sponsored the project and both he and Doug McIlwraith assisted in the drafting of the manuscript. All authors gave final approval for publication.}

\competing{We declare we have no competing interests.}

\funding{No directly attributable sources of funding were used for this research.}

\ack{David Birch provided valuable visualization assistance in Imperial's Data Observatory.  This research was carried out in the Imperial College-Zhejiang University Joint Laboratory for Applied Data Science at the Imperial College Data Science Institute.}

\disclaimer{Nothing to disclaim.}


\bibliographystyle{RS}
\bibliography{99_references}
\end{document}